\documentclass[twocolumn]{aastex63}
\usepackage[utf8]{inputenc}
\usepackage{comment}

\begin{document}
\title{A more probable explanation for a continuum flash in the direction of a redshift $\approx$ 11 galaxy} 


\author[0000-0003-3780-6801]{Charles L. Steinhardt}
\affiliation{Cosmic Dawn Center (DAWN)}
\affiliation{Niels Bohr Institute, University of Copenhagen, Jagtvej 128, K\o benhavn N, DK-2200, Denmark}
\author[0000-0002-8109-033X]{Michael I. Andersen}
\affiliation{Cosmic Dawn Center (DAWN)}
\affiliation{Niels Bohr Institute, University of Copenhagen, Jagtvej 128, K\o benhavn N, DK-2200, Denmark}
\author[0000-0003-2680-005X]{Gabriel B. Brammer}
\affiliation{Cosmic Dawn Center (DAWN)}
\affiliation{Niels Bohr Institute, University of Copenhagen, Jagtvej 128, K\o benhavn N, DK-2200, Denmark}
\author[0000-0001-8415-7547]{Lise Christensen}
\affiliation{Cosmic Dawn Center (DAWN)}
\affiliation{Niels Bohr Institute, University of Copenhagen, Jagtvej 128, K\o benhavn N, DK-2200, Denmark}
\author[0000-0002-8149-8298]{Johan P. U. Fynbo}
\affiliation{Cosmic Dawn Center (DAWN)}
\affiliation{Niels Bohr Institute, University of Copenhagen, Jagtvej 128, K\o benhavn N, DK-2200, Denmark}
\author[0000-0002-2281-2785]{Bo Milvang-Jensen}
\affiliation{Cosmic Dawn Center (DAWN)}
\affiliation{Niels Bohr Institute, University of Copenhagen, Jagtvej 128, K\o benhavn N, DK-2200, Denmark}
\author[0000-0001-5851-6649]{Pascal A. Oesch}
\affiliation{Cosmic Dawn Center (DAWN)}
\affiliation{Niels Bohr Institute, University of Copenhagen, Jagtvej 128, K\o benhavn N, DK-2200, Denmark}
\affiliation{University of Geneva, Department of Astronomy, Chemin Pegasi 51, 1290 Versoix, Switzerland}
\author[0000-0003-3631-7176]{Sune Toft}
\affiliation{Cosmic Dawn Center (DAWN)}
\affiliation{Niels Bohr Institute, University of Copenhagen, Jagtvej 128, K\o benhavn N, DK-2200, Denmark}

\begin{abstract}
Recent work reported the discovery of a gamma-ray burst (GRB) associated with the galaxy GN-z11 at $z\sim11$. The extreme improbability of the transient source being a GRB in the very early Universe requires robust elimination of all plausible alternative hypotheses.  We identify numerous examples of similar transient signals in separate archival MOSFIRE observations and argue that Solar system objects --- natural or artificial --- are a far more probable explanation for these phenomena.  An appendix has been added in response to additional points raised in Jiang et al. (2021), which do not change the conclusion.
\end{abstract} 





\section{Introduction}

\citet{Jiang2020} recently reported a transient source  in MOSFIRE \citep{MOSFIRE} $K$ band slit spectroscopy of GN-z11 \citep{Oesch2016}, the highest-redshift galaxy yet observed.  They interpret this transient as the possible discovery of a UV flash from a gamma-ray burst (GRB) at $z = 10.957$.  However, as shown in \S~\ref{sec:MOSFIRE}, a MOSFIRE detection of a Solar system object or spacecraft is a far more likely explanation.

\section{Transient Signals from Solar System Objects in MOSFIRE Spectra}
\label{sec:MOSFIRE}

\begin{figure*}
    \centering
    \includegraphics[width=1\textwidth]{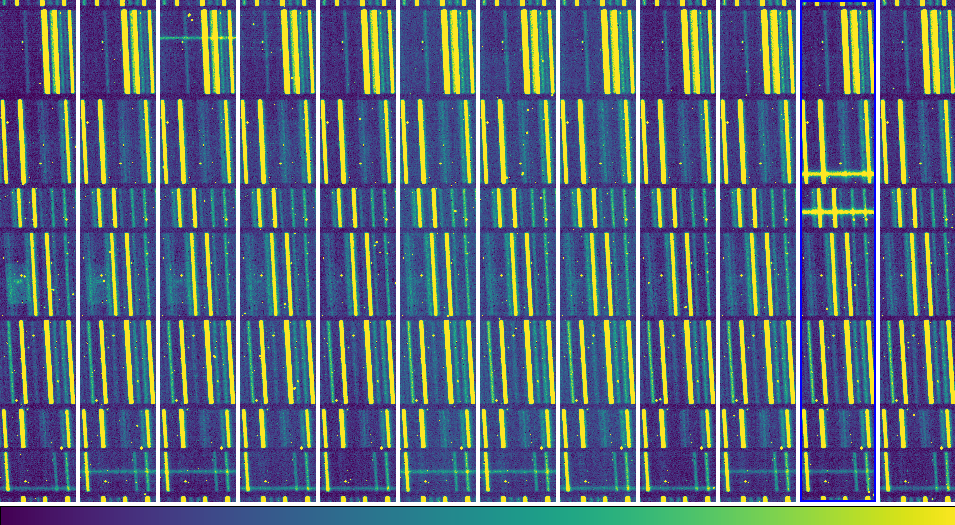}
    \caption{Twelve raw MOSFIRE spectra of an Abell 1689 field taken in sequence, with each spectrum having an exposure time of 120 seconds. Only a smaller part of each spectrum is shown. A ``flash'', i.e.\ a trace that only appears in a single exposure, is seen in the 3rd exposure (\texttt{MF.20170307.46851.fits}) in the top slit, and a double flash is seen in the 11th exposure (\texttt{MF.20170307.48096.fits}) in the two slits below.}
    \label{fig:A1689-2D-12exposures}
\end{figure*}

\begin{figure}
    \centering
    \includegraphics[width=0.5\textwidth]{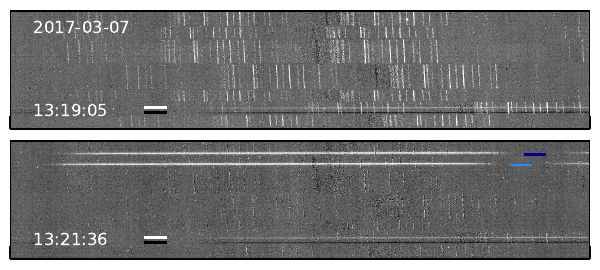}
    \caption{A double ``flash'' seen by MOSFIRE on 7 Mar 2017 while observing the Abell 1689 field.
    Shown are two sequential $J$ band difference images of the $B-A$ nod positions where the B exposures were taken just 2.5 minutes apart.  The white and black bars indicate the characteristic positive and negative spectra of a bright alignment star seen in both difference images.  The flashes indicated by the blue bars appear in a single 120~s exposure.}
    \label{fig:A1689-2D}
\end{figure}

\begin{figure*}
    \centering
    \includegraphics[width=1\textwidth]{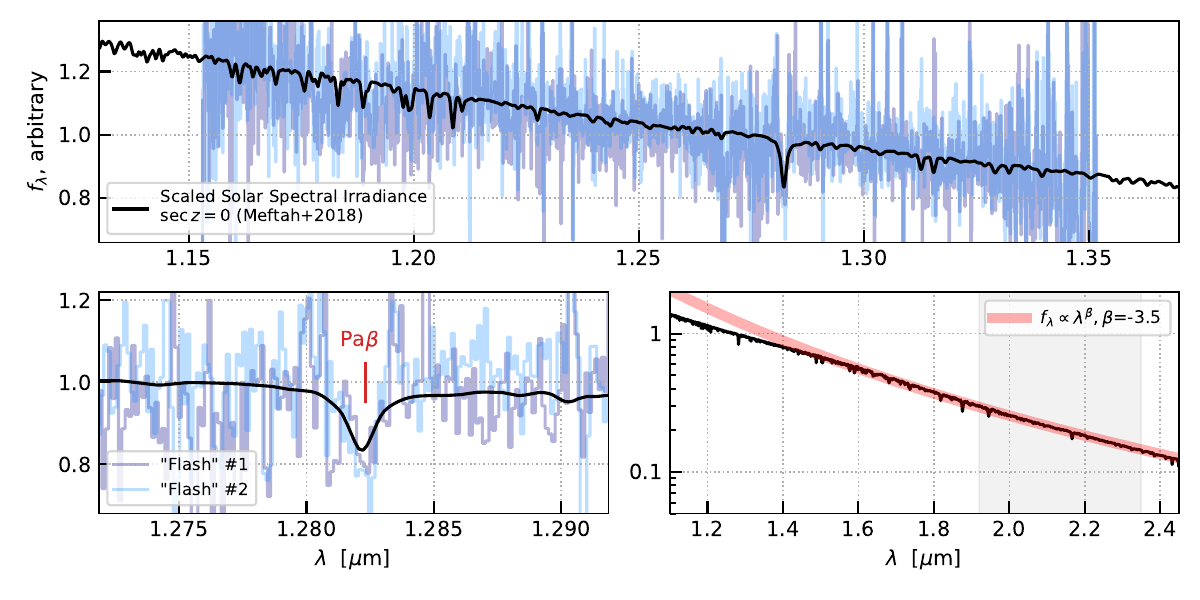}
    \caption{\textit{Top:} Telluric-corrected $J$ band spectra of the two flashes shown in Fig.~\ref{fig:A1689-2D}.  The black line shows the Solar spectrum from \cite{Meftah2018} normalized to the flash spectra. \textit{Lower left:} The flash spectra show an absorption feature consistent with the wavelength and depth of the Paschen-$\beta$ line in the Solar spectrum.  \textit{Lower right:} Expanded near-infrared view of the Solar spectrum, which has a power-law slope $\beta\approx-3.5$ (red curve) in the $K$ band (shaded grey region).}
    \label{fig:A1689-1D}
\end{figure*}

To date\footnote{\url{https://koa.ipac.caltech.edu/UserGuide/mosfire.csv}} MOSFIRE has obtained 81,776 spectroscopic exposures of science targets.  If \textit{any} other exposures show similar transients as that reported for GN-z11, then the chance probability of finding such a transient in a single random exposure, regardless of its origin, is $>$$10^4$ times more likely than a serendipitous GRB detection with probability $\ll 10^{-8}$.  In a visual search of 12,300 exposures from the \textit{Keck} MOSFIRE archive we find a minimum of 27 single-exposure transients in the files summarized in Table~\ref{table:exposures}.  

\begin{figure*}
    \centering
    \includegraphics[width=0.75\textwidth]{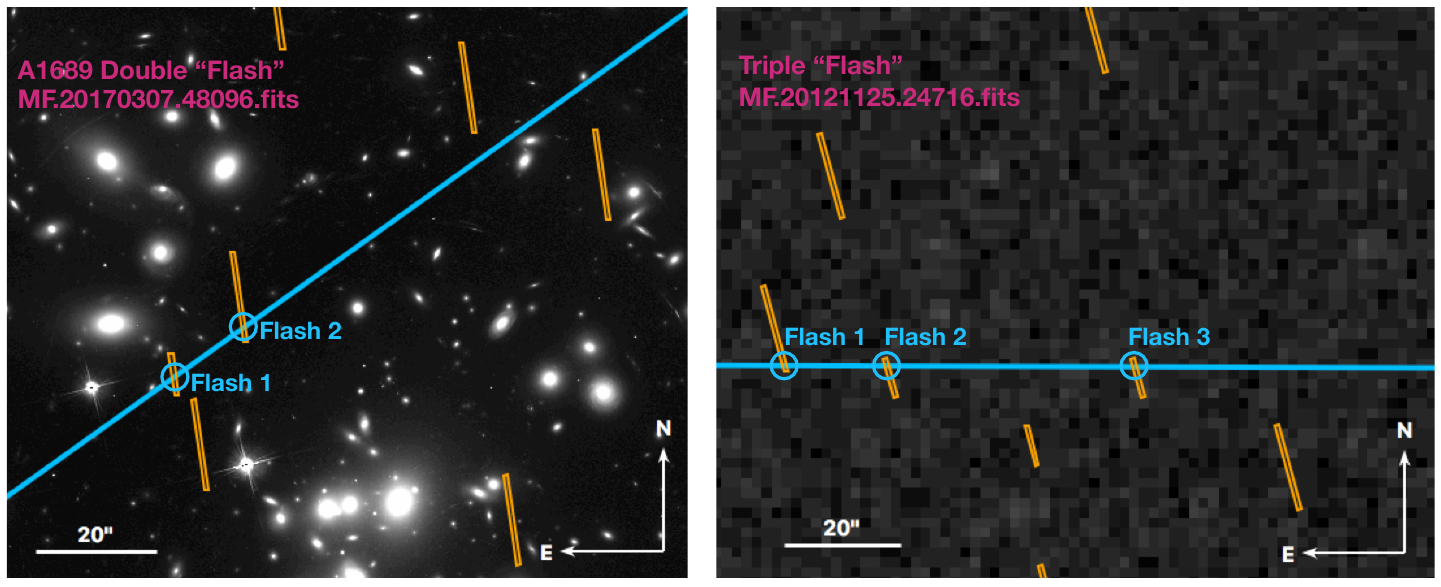}
    \caption{Slit mask layouts for two of the MOSFIRE frames with multiple detected flashes in one exposure. The position of these flashes (blue circles) is consistent with a source moving on a straight trajectory that crosses multiple slits in a single exposure (blue line). This strongly suggests that such transients occur due to Earth-orbiting satellites often seen in imaging observations.}
    \label{fig:FlashTrajectories}
\end{figure*}

Three of these transients appear in a sequence of 12 exposures (Fig.~\ref{fig:A1689-2D-12exposures}) taken on 7 Mar 2017 centered on the lensing cluster Abell 1689 (A1689), and two of those are seen in a single 120~s exposure (Fig.~\ref{fig:A1689-2D}).  As with the GN-z11 flash, this pair of A1689 transient spectra shows telluric absorption features indicating their origin above the atmosphere.  Both transients in the pair have the same spectrum (Fig.~\ref{fig:A1689-1D}) suggesting that they arise from a single source moving with a minimum angular speed $>0\farcs1$ per second that passed through both slits (Fig.~\ref{fig:FlashTrajectories}).  Furthermore, the shapes of their spectra are fully consistent with reflected Solar spectra, both in the overall slope and also with the detection of the Paschen-$\beta$ absorption feature.  

The steeper slope of the GN-z11 transient in the $K$ band is also consistent with a reflected Solar spectrum (Fig.~\ref{fig:A1689-1D}).  It is a featureless $K$ band spectrum consistent with a powerlaw $f_{\lambda}\propto \lambda^{-3.2\pm0.4}$, which is significantly steeper than regularly found for lower-redshift GRB afterglows that have observed UV-to-optical slopes in the range between $-0.9$ and $-1.6$ \citep{Jakobsson2004,Japlj2015,Li2015}, and indeed \citet{Jiang2020} state that the transient is not caused by a regular GRB reverse shock emission.  In addition to the non-thermal GRB afterglow spectra, some GRBs exhibit a prompt thermal blackbody emission presumably from a shock-breakout \citep{Campana2006}. For a Pop III GRB, the dominant contribution from the prompt thermal emission component lies between $10^2\text{--}10^5$ keV in the rest-frame \citep{Toma2011}, i.e. the thermal component peaks in X-ray emission. \citet{Jiang2020} argued that a slope of $-3.2$ can arise in a self absorbed synchrotron spectrum, but this requires fine-tuning of the model parameters of \citet{Uhm2014}.  A reflected Solar spectrum is a better fit, as it does not require this fine-tuning.

It was not possible to identify any of the transient MOSFIRE sources with specific Solar system objects in the NASA HORIZONS database\footnote{\url{https://ssd.jpl.nasa.gov/?horizons}} of natural bodies (e.g., asteroids, comets) and select spacecraft.  Based on the spectral similarity to the reflected Solar spectrum and the angular speed of the moving sources shown in Fig.~\ref{fig:FlashTrajectories}, the transients detected in MOSFIRE are most likely high-altitude Earth-orbiting satellites.  

\citet{Jiang2020} argue that the probability of such an interloper is very low based on the telescope pointing and assumed distribution of satellite orbits.  However, 8 additional transients were found among $\sim$6000 archival exposures of the same GOODS-North field.  The probability of finding one in a specific slit is perhaps 10-20 times lower, although the other arguments for implausibility would have applied to a transient in any slit in the mask.  Earth-orbiting satellites cause trails that are so ubiquitous in \textit{Hubble Space Telescope} images seen from low-Earth orbit as to motivate the development of automated software to detect and remove them \citep{Borncamp2019}---to say nothing of the constellations of low-Earth orbit satellites now regularly photo-bombing both professional and amateur images from the ground \citep{Tyson2020}.  Identifying the specific satellites that cause the trails in the MOSFIRE exposures summarized in Table~\ref{table:exposures} is beyond the scope of this work, and the \texttt{calsky} online database \cite{Jiang2020} used to search for satellite counterparts to the GN-z11 transient is no longer available to make that specific comparison.  

\citet{Jiang2020} also argue that time of day and compactness of the GN-z11 flash are impossible for a man-made satellite.  However, examples exist with both properties.  The exposure \texttt{MF.20140425.29736.fits}, in GOODS-North, shows an H band transient obtained at UT 08:15 (the GN-z11 transient was observed at UT 08:07).  The NAVSTAR 52 (USA 168, NORAD ID 27704) satellite, with an orbital inclination angle of 54 degrees, is in the approximate vicinity of the GN-z11 flash on the night of observations, with a direction of travel close to east-west.  A satellite with a similar orbit could have produced a compact flash like the one reported. 

\section{Discussion}

The goal of this analysis is to assign the most probable explanation to the possible UV flash reported by \citet{Jiang2020} in GN-z11.  In terms of brightness the observations are not incompatible with being related to a GRB \citep{Kann2020}.  However, the spectral shape is significantly bluer than what is typical for GRB afterglows, and more similar to those shown in Fig. \ref{fig:A1689-2D} in MOSFIRE observations of Abell 1689, likely natural Solar system objects or spacecraft.   

As estimated in \citet{Jiang2020}, the probability merely of detecting a GRB is between $\sim 10^{-10}$ and $10^{-8}$ under $\Lambda$CDM, and the GN-z11 flash does not have the SED of a typical GRB.  Although \citet{Kann2020} argue that such an SED is plausible, given current catalogs, the probability of finding one is $\lesssim 10^{-4}$ less likely than merely finding a GRB.  So, the probability of detecting a GRB of this type is $\lesssim 10^{-13}$.  

$10^{-13}$ is small enough that many other improbable explanations are far more likely.  From an archival search of MOSFIRE observations, the rate of natural bodies and/or spacecraft causing flashes in dispersed spectra is of order $10^{-3}$ per exposure, making this explanation $\gtrsim 10^{10}$ times more likely.  Due to the rate of these interlopers at other observatories, even \emph{in the absence of these examples}, this was the most probable explanation.  At least three of these transients, including the one in the GN-z11 slit, have spectra indistinguishable from a (reflected) Solar spectrum, further supporting Earth-orbiting satellites as their most probable cause.

\acknowledgements 

CS is supported by ERC grant 648179 "ConTExt".  The Cosmic Dawn Center (DAWN) is funded by the Danish National Research Foundation under grant No.\ 140. BMJ is supported in part by Independent Research Fund Denmark grant DFF - 7014-00017. JPUF acknowledges support from the Carlsberg foundation. This research has made use of the Keck Observatory Archive (KOA), which is operated by the W. M. Keck Observatory and the NASA Exoplanet Science Institute (NExScI), under contract with the National Aeronautics and Space Administration.  The authors wish to recognize and acknowledge the very significant cultural role and reverence that the summit of Maunakea has always had within the indigenous Hawaiian community.  We are most fortunate to have the opportunity to conduct observations from this mountain.

\begin{table}[ht]
\begin{center}
\begin{tabular}{lcc}
Keck Archive Filename & Grating & UT\\
\hline                         
\hline                         
GOODS-North &  & \\                 
\hline                         
\texttt{MF.20130114.54832.fits} & K & 15:13\\
\texttt{MF.20130215.56333.fits} & K & 15:38\\
\texttt{MF.20130322.54354.fits} & H & 15:05\\
\texttt{MF.20140425.29736.fits} & H & 08:15\\
\texttt{MF.20150110.55808.fits} & K & 15:30\\
\texttt{MF.20160102.57635.fits} & H & 16:00\\
\texttt{MF.20160128.57006.fits} & H & 15:50\\
\texttt{MF.20170304.45898.fits} & J & 12:44\\
\texttt{MF.20170407.29239.fits}\tablenotemark{$\dagger$} & K & 08:07 \\
\hline                          
\hline                          
Other &  & \\                      
\hline                        
\texttt{MF.20121012.46713.fits} & K & 12:58 \\
\texttt{MF.20121125.24716.fits}\tablenotemark{$\ast$} & H & 06:51\\
\texttt{MF.20121206.49727.fits} & H & 13:48 \\
\texttt{MF.20130104.31845.fits} & K & 08:50 \\
\texttt{MF.20131128.45592.fits} & K & 12:39 \\
\texttt{MF.20131224.25147.fits} & J & 06:59 \\
\texttt{MF.20131224.34612.fits} & J & 09:36 \\
\texttt{MF.20170304.24709.fits} & J & 06:51 \\
\texttt{MF.20170307.46851.fits} & J & 13:00 \\
\texttt{MF.20170307.48096.fits}\tablenotemark{$\star$} & J  & 13:21 \\
\texttt{MF.20170416.25674.fits} & H & 07:07 \\
\texttt{MF.20170416.51654.fits} & H & 14:20 \\
\texttt{MF.20170507.34054.fits} & J & 09:27 \\
\texttt{MF.20170508.41251.fits} & J & 11:27 \\
\texttt{MF.20170508.41560.fits} & J & 11:32 \\
\texttt{MF.20170508.41712.fits} & J & 11:35 \\
\texttt{MF.20170508.42020.fits} & J & 11:40 \\
\texttt{MF.20170508.49166.fits} & J & 13:39  
\end{tabular}
\caption{Keck archive filenames of MOSFIRE exposures with identified transient flashes.  These files can be downloaded directly from the Keck Observatory Archive at URLs like \href{https://koa.ipac.caltech.edu/cgi-bin/getKOA/nph-getKOA?filehand=/koadata34/MOSFIRE/20130114/lev0/MF.20130114.54832.fits}{MF.20130114.54832.fits}.  Notes: $\dagger$ GN-z11, \cite{Jiang2020}; $\ast$ Triple flash, Fig.~\ref{fig:FlashTrajectories}; $\star$ Double flash, Figs.~\ref{fig:A1689-2D-12exposures} - \ref{fig:FlashTrajectories}}
\label{table:exposures}
\end{center}
\end{table}

\bibliographystyle{mnras}
\bibliography{refs.bib} 

\appendix

Following the initial posting of this work, \citet{Jiang2021} raised several additional points, which are addressed here.  It should be noted that Jiang et al. use these arguments to bound the probability of an interloper spacecraft at $3 \times 10^{-7}$.  For comparison, \citet{Jiang2020} estimate the corresponding probability of GRB detection as $(0.3 - 60) \times 10^{-10}$ (this work estimates the probability as $\lesssim 10^{-13}$).  Thus, the results in \citet{Jiang2021} are in agreement with this paper that a spacecraft is the far more probable explanation.

However, we disagree with several of the arguments for a lower spacecraft probability raised in Jiang et al. (2021).  Jiang et al. argue that the GN-z11 flash is special because it coincides with a known object.  However, MOSFIRE slit masks are always constructed to take spectra of known objects; in that sense, every object in which a flash is found would be special.  More generally, this is a version of the well-known multiple comparison problem \citep{Tukey1949,Kramer1956,Sidak1967,Holm1979}.

They also argue that the estimate in this work must be flawed, since otherwise there should be approximately one spacecraft per 5 square degrees.  This is a slight overestimate, because a single spacecraft can generate multiple flashes within a single observation (Fig. \ref{fig:FlashTrajectories}).  However, with approximately tracked 5400 spacecraft with orbit perigees higher than 1000~km\footnote{\url{space-track.org}}, one per 5 square degrees is reasonable to within an order of magnitude and the number of artificial objects that could cause the types of transients discussed here is almost certainly much higher still (e.g., untracked debris in high geosynchronous orbits; \citealt{Corbett2020, Blake2021, Nir2021}).

\begin{figure*}
    \centering
    \includegraphics[width=0.95\textwidth]{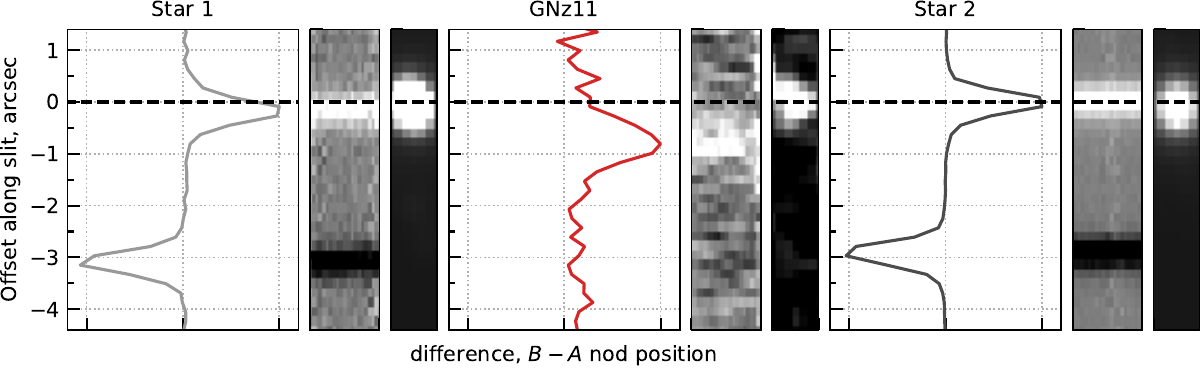}
    \caption{Source alignment within the MOSFIRE slits.  Three slits are shown in groups of three panels: an alignment star in the second slit from the top of the detector (``S-29371''; {\it left}), GN-z11 ({\it center}) and a bright alignment star in the slit just below GN-z11 (``S-17541''; {\it right}).  In each case, the curve shows the average cross-dispersion profile of the difference image created from the exposure containing the transient (nod position ``B'') and the exposure preceding it (position ``A'').  A slice of the 2D spectroscopic difference image is also shown, and the third, thinnest panel shows the \textit{Hubble} F160W image at the same position, convolved with a 2D Gaussian to approximate the ground-based seeing.  The transient in the GN-z11 slit is offset from the galaxy center by $0\farcs7$.  It is also more extended than the two point sources by a factor of $\sim$2.}
    \label{fig:slit_alignment}
\end{figure*}

Finally, Jiang et al. argue that the flash must be associated with GN-z11 because of its proximity to the center of the galaxy.  A careful analysis of the target alignment within the slits finds that the traces of two alignment stars in slits separated by half the instrumental field of view are well aligned with the known target positions while the transient trace is clearly offset from the galaxy center by $\sim0\farcs7$ (Fig.~\ref{fig:slit_alignment}), which would correspond to 2.8~kpc at $z=11$.  The two stars demonstrate the reliability of the absolute astrometry, and ``Star 2'' in the slit just below GN-z11 provides an unambiguous relative measurement of the offset.  As described by \cite{Jiang2020}, the transient is nearly twice as extended as the two point sources.  Although the offset and source extent aren't dispositive of the GRB interpretation of the transient, they would require yet another coincidence to support it.  For example, the GRB probability argument would need to account for the dramatically lower stellar density at nearly 5 half-light radii \citep[$R_\mathrm{e}=0.6\pm0.3~\mathrm{kpc}$;][]{Oesch2016}.

As a result, Jiang et al. (2020) argue that it is not merely improbable, but rather impossible for such a flash to have come from a spacecraft.  So, the offset would not be relevant, because even a non-extended flash near the edge of the slit would have no other possible explanation, and thus must be a GRB.  This impossibility argument is essential, because any estimate of the expected spacecraft rate, including the estimate in Jiang et al. (2021), concludes that it is far higher than the expected GRB rate. 

Jiang et al. argue that because some (although not all) of the flashes in Table \ref{table:exposures} differ from GN-z11 in observation time or required orbital inclination angle, GN-z11 must be a unique object with a different explanation.  However, the examples in Table \ref{table:exposures} include a similar flash in GOODS-N at a similar time of night, and this work identified an example of a satellite on an orbit which would produce a non-extended flash in similar observations.  

There are certainly many cases in the history of astronomy when what was originally thought to be one class of objects instead turned out to be a combination of objects with several astrophysical origins.  However, if the Jiang et al. argument were correct, it would apply to several similar flashes found in the MOSFIRE archive.  At such a rate, MOSFIRE would be a more efficient GRB detector than \emph{Swift}.  The preferred explanation should instead be the most probable one, which both Jiang et al. (2021) and this work agree is a spacecraft.

\end{document}